# Measuring field-normalized impact of papers on specific societal groups: An altmetrics study based on Mendeley data

Lutz Bornmann[*] and Robin Haunschild[**]

[*]Corresponding author:

Division for Science and Innovation Studies

Administrative Headquarters of the Max Planck Society

Hofgartenstr. 8,

80539 Munich, Germany.

Email: bornmann@gv.mpg.de

[**] Contributing author:

Max Planck Institute for Solid State Research

Heisenbergstr. 1,

70569 Stuttgart, Germany.

Email: R.Haunschild@fkf.mpg.de

**Abstract**

Bibliometrics is successful in measuring impact, because the target is clearly defined: the publishing scientist who is still active and working. Thus, citations are a target-oriented metric which measures impact on science. In contrast, societal impact measurements based on altmetrics are as a rule intended to measure impact in a broad sense on all areas of society (e.g. science, culture, politics, and economics). This tendency is especially reflected in the efforts to design composite indicators (e.g. the Altmetric attention score). We deem appropriate that not only the impact measurement using citations is target-oriented (citations measure the impact of papers on scientists), but also the measurement of impact using altmetrics. Impact measurements only make sense, if the target group – the recipient of academic papers – is clearly defined. Thus, we extend in this study the field-normalized reader impact indicator proposed by us in an earlier study, which is based on Mendeley data (the mean normalized reader score, MNRS), to a target-oriented field-normalized impact indicator (e.g., $MNRS_{ED}$ measures reader impact on the sector of educational donation, i.e., teaching). This indicator can show – as demonstrated in empirical examples – the ability of journals, countries, and academic institutions to publish papers which are below or above the average impact of papers on a specific sector in society (e.g., the educational or teaching sector). For example, the method allows to measure the impact of scientific papers on students – controlling for the field in which the papers have been published and their publication year.

# 1    Introduction

Numbers are absolutely essential when it comes to different kinds of judgments (e.g., of research groups or institutions) in science (Marewski & Bornmann, in preparation). One of the most frequently used sources for numbers are publication and citation data from bibliometrics. The term altmetrics (the abbreviation of alternative metrics) was created by Jason Priem who introduced it in the scientometric community in the altmetrics manifesto (Priem, Taraborelli, Groth, & Neylon, 2010). According to Wilsdon et al. (2015) the term altmetrics "is variously used to mean 'alternative metrics' or 'article level metrics', and it encompasses webometrics, or cybermetrics, which measure the features and relationships of online items, such as websites and log files. The rise of new social media has created an additional stream of work under the label altmetrics. These are indicators derived from social websites, such as Twitter, Academia.edu, Mendeley, and ResearchGate with data that can be gathered automatically by computer programs" (pp. 5-6).

At least two developments lead to the interest of scientometricians and the consumers of their studies in altmetrics: (1) In the context of post-academic science with its focus on accountability, governments and funders are interested in measurements of returns on investment from research, i.e. in measurements of economic benefits of publicly funded research (Mohammadi, Thelwall, & Kousha, 2016; Ziman, 2000). Altmetrics have been proposed as a possibility to measure impact of research on sections of society (Darling, Shiffman, Côté, & Drew, 2013; Weller, 2015); a convincing empirical validation of this proposal is still missing though. For Wilsdon et al. (2015) "the systematic use of alternative indicators as pure indicators of academic quality seems unlikely at the current time, though they have the potential to provide an alternative perspective on research dissemination, reach and 'impact' in its broadest sense" (pp. 45-46). (2) There is a swift observable in recent years from print to online in the publication process as well as in the reading and discussion of publications. Early studies in webometrics (Bollen, Van de Sompel, Hagberg, & Chute, 2009) revealed the richness of the digital patterns for scientometric studies (Adie, 2014).

The sources of altmetrics are usually social media platforms, but they are not restricted to these platforms (Work, Haustein, Bowman, & Larivière, 2015). Typical sources are, for example, Twitter, Mendeley, Facebook, F1000, policy documents, blogs. In altmetrics, views, downloads, clicks, notes, saves, tweets, shares, likes, tags, posts, recommends, trackbacks, mentions, discussions, bookmarks, and comments are counted from these and other sources (Bornmann, 2014a). Wilsdon et al. (2015) considers comments, recommends, bookmarks, and tweets of especial importance. Today, altmetrics are used by publishers (e.g. Wiley, Springer, or BioMed Central) which add altmetrics to their papers (Thelwall & Kousha, 2015). Altmetrics are provided by companies, like ImpactStory, Plum Analytics, and Altmetric, and recommended by Snowball Metrics Recipe Book (Colledge, 2014). In this book, some universities (in cooperation with Elsevier) try to standardize their way in measuring institutional output and impact.

This study focusses on Mendeley – one of the most important online reference managers which data is frequently used for altmetrics studies – to propose a method of measuring impact field-normalized on specific status groups. For example, the method allows to measure the impact of scientific papers on students – controlling for the field in which the papers have been published and their publication year.

## 2     Research questions

Bibliometrics is successful in measuring impact, because the target is clearly defined: the publishing scientist who is still active and working. Thus, citations is a target-oriented metric which measures impact on science (Lähteenmäki-Smith, Hyytinen, Kutinlahti, & Konttinen, 2006). In contrast, societal impact measurements based on altmetrics are as a rule intended to measure impact in a broad sense on all areas of society (e.g. science, culture, politics, and economics). For example, Mendeley data are used to measure the broad impact of research published by institutions (Colledge, 2014). Furthermore, there is the tendency in broader impact measurements based on altmetrics to use different altmetrics sources for the calculation of a composite indicator. These indicators have been developed because many altmetrics (e.g. Facebook or blog counts) are characterized by low numbers. For example, Altmetric has introduced the Altmetric Attention Score. The score measures the impact (attention) of publications over very different altmetrics sources, which are frequently not comparable (e.g., blog posts and tweets). Each source is scaled with a rather arbitrary factor.[1]

We deem appropriate that not only the impact measurement using citations is target-oriented (citations measure the impact of papers on scientists), but also the measurement of impact using altmetrics. Altmetric and Scholastica (2015) give the following relevant examples here: "Someone publishing a study on water use in Africa may be particularly keen to see that many of those tweeting and sharing the work are based in that region, whereas economics scholars might want to keep track of where their work is being referenced in public policy or by leading think-tanks" (p. 24). Another example is presented by Roemer and Borchardt (2015) focusing on conferences: "From this perspective, it's no longer enough to say, 'I presented at a conference!' Instead, researchers must use measurements like audience size, presentation feedback, and Twitter mentions to provide evidence that others found their presentation meaningful in some way" (p. 18). Without the restriction of impact measurements to specific target areas (e.g. science, education), these measurements do not make sense in the research evaluation context: Whereas citations measure impact on science, it is not clear which kind of impact data from altmetrics sources measure. The general uncertainty in scientometrics (and beyond) about the meaning of altmetrics is probably based on the tendency to use composite indicators (e.g. the Altmetric attention score) or counts without target-restrictions when altmetrics are used.

---

[1] https://help.altmetric.com/support/solutions/articles/6000060969-how-is-the-altmetric-attention-score-calculated-

Mendeley which was bought by Elsevier in April 2013 (Rodgers & Barbrow, 2013) is a website (http://mendeley.com) and a program which allows registered users to enter information about publications of interest. The saved publication information can be used to create reference lists for academic works (e.g. bachelor or master thesis) and can be shared with others (Thelwall & Kousha, 2015). Used as altmetrics, data from Mendeley show how often publications are saved by users which might be an indication of readership. The survey of 860 Mendeley users by Mohammadi et al. (2016) shows that "it is reasonable to use Mendeley bookmarking counts as an indication of readership because most (55%) users with a Mendeley library had read or intended to read at least half of their bookmarked publications. This was true across all broad areas of scholarship except for the arts and humanities (42%)."

According to Wouters et al. (2015) "Mendeley readership bookmarks seem to be the most promising altmetric indicator" (p. 89). For Haustein (2014), "Mendeley may be the most promising new source for evaluation purposes as it has the largest user population, the greatest coverage, highest number of readers per document, and strongest correlations between usage and citation counts." Mendeley data are especially suitable as source for calculating (target-oriented) metrics, because the coverage of publications is generally high (compared to other altmetrics). This has been demonstrated by several studies as the overviews by Work et al. (2015) and Thelwall and Kousha (2015) show. Since Mendeley data also allow the differentiation in self-assigned professional status groups (e.g. students, professors, or researchers), Mendeley is particularly interesting in (broader) impact measurements target-oriented. For example, one can measure the impact of publications on the third level education – if the measurement of impact is restricted to students.

This kind of target-oriented measurements will be explained in this study: In concrete terms, the field-normalized reader impact indicator proposed by us (Haunschild & Bornmann, 2016a) based on Mendeley data is extended here to a target-oriented field-normalized impact indicator. This indicator can show – as demonstrated in section 5.4 – the ability of academic institutions to publish papers which are below or above the average impact of papers on the educational sector.

## 3      Literature overview on Mendeley studies

Mendeley is a reference management tool for people organizing, sharing and discovering research (Gunn, 2013). The high number of Mendeley members points out that this tool is more attractive than other tools (e.g., CiteULike) for people working with publications (Kumar Das & Mishra, submitted; Van Noorden, 2014). The attractiveness among people working with published material is also an important reason, why the tool is one of the most frequently used sources for altmetrics. In all likelihood, "more frequently saved articles are likely to be more frequently read" (Li, Thelwall, & Giustini, 2012, p. 464). Further points for its attractiveness are that (1) the publications in Mendeley are not limited by the indexes of bibliographic databases (e.g. Web of Science, Thomson Reuters, or Scopus, Elsevier), (2) readers do not have to be authors in order to count their reading (as it is the case with citations) (Li et al., 2012), (3) readership data are generated as by-products of workflows (in journalism, teaching, or learning) (Haustein, 2014), and (4) readership of publications can be counted relatively fast after publication, typically they only take a few weeks or months to accumulate (Lin & Fenner, 2013).

According to Mohammadi and Thelwall (2014) the data from social bookmarking tools may "help to overcome the lack of global and publisher-independent usage data." However, the data should be handled with care, because posting publications on Mendeley can have various motives: "Other people might be interested in this paper. I might read this paper in the future. I have read this paper and want it to be easily findable. I want other people to think I have read this paper. It is my paper, and I maintain my own library. It is my paper, and I want people to read it. It is my paper, and I want people to see that I wrote it. I might skim read this paper in the future because I suspect it might back up an argument I'm thinking about making and it looks like it would make a useful citation" (Taylor, 2013, p. 20). Further critical points in dealing with Mendeley data are that the user-generated data are not quality controlled and can be very easily spammed by users or fake Mendeley profiles (Thelwall & Kousha, 2015).

Most of the altmetric studies which have analysed Mendeley data focussed on two topics: (1) coverage of publications in Mendeley and (2) correlations between reader and citation counts.

Ad 1) In a comprehensive study, Haustein and Larivière (2014a) analysed more than one million papers published in journals covered by both PubMed and WoS. They published a detailed table with proportions of papers having at least one reader in several disciplines. The results show, e.g., with 79% a high proportion in biophysics, but a low proportion in urology. Furthermore, there is a high coverage of publications in many sub-disciplines of psychology. Similar results have been reported from the analysis of 20,000 random publications from the WoS (Zahedi, Costas, & Wouters, 2014b), documents authored by presenters of the 2010 STI conference in Leiden (Haustein, Peters, et al., 2014), and nearly 500 papers published in *Aslib proceedings* (Haustein & Larivière, 2014b). In these studies, the proportion of publications with at least one reader in Mendeley is 63% (Zahedi et al., 2014b), 82% (Haustein, Peters, et al., 2014), and 77% (Haustein & Larivière, 2014b). For Priem (2014), the good coverage of publications on Mendeley gives birth to a rival of commercial databases like WoS and Scopus.

Ad 2) A comparable high number of publications have dealt with the analysis of correlations between readership and citation counts. For example, Weller and Peters (2012) considered nearly 1000 publications from 41 authors in their analysis and found a correlation coefficient of approximately 0.5 between readership and citation counts. For *Nature* and *Science* publications, the correlation coefficients are similarly high (Li et al., 2012). Literature overviews on studies reporting correlation coefficients can be found in Bar-Ilan, Shema, and Thelwall (2014) and Haustein, Peters, et al. (2014). The meta-analysis of Bornmann (2015a) tries to formulate a general and summarizing statement on the correlation between readership and citation counts. Meta-analysis is a statistical approach that combines evidence from many different studies to obtain an overall estimate of treatment effects. The results show that the pooled coefficient over several altmetric studies is approximately 0.5. With this result, readership counts are correlated stronger with citation counts than other altmetrics (e.g., Twitter and blog counts). For Priem (2014), the moderate correlation between readership and citation counts "is likely because of the importance of reference managers in the citation workflow. However, the lack of perfect or even strong correlation suggests that this altmetric, too, captures influence not reflected in the citation record."

The proposal in this study not to use the complete set of readers, but to focus on specific reader groups, can be seen as an important step to differentiate in a better way between citation impact and reader impact. Target-oriented reader impact can focus on the research sector (e.g., by including impact on professors) or on the educational sector (e.g. by including impact on Bachelor students).

## 4 Datasets used

We retrieved 1,493,838 papers (1,416,284 article and 77,554 reviews) published in 2014 from our Web of Science (WoS) in-house database – derived from the Science Citation Index Expanded (SCI-E), Social Sciences Citation Index (SSCI), and Arts and Humanities Citation Index (AHCI) provided by Thomson Reuters (Philadelphia, USA). For 95.33% (n = 1,350,634) of the papers, a DOI was available in the WoS database. We searched for the DOIs of the remaining papers in the CrossRef database using the Perl module Bib::CrossRef.[2] The API returned a DOI for each paper. Therefore, we checked if the publication year, journal title and issue agree with the values in our in-house database. In 105,409 of the cases, the bibliographic values disagreed and the papers were discarded. Using this procedure, we obtained 37,792 (2.53%) additional DOIs.

---
[2] See http://search.cpan.org/dist/Bib-CrossRef/lib/Bib/CrossRef.pm

In total, we retrieved the Mendeley reader statistics for 1,388,426 papers via the DOI from the Mendeley API made available in 2014. The Mendeley data were gathered between July 3 and July 8, 2016. We used R (R Core Team, 2014) to interface to the Mendeley API. We found 1,239,445 of the papers (89.27%) at Mendeley. The remaining 148,982 papers were treated as papers without Mendeley readers. In addition, we found 834 papers at Mendeley registered with a reader count of zero. Altogether, 149,816 papers (10.79%) have a reader count of zero. Overall, 16,950,398 reader counts were observed for all papers (12.21 reader counts per paper on average). The academic status information (UK and US biased) is available for 16,949,442 of the bookmarks (reads) by Mendeley users (99.99%).

## 5 Results

### 5.1 Differences in reader counts between status groups

Earlier, we have empirically demonstrated that there are significant differences in reader data between WoS categories (Bornmann & Haunschild, 2016b; Haunschild & Bornmann, 2016a). For reviews published in 2012, e.g., we found with 85.22 the highest average number of readers in "Psychology, Experimental" and with 0.27 the lowest (non-zero) number in "Literature" (Haunschild & Bornmann, 2016a). Against the backdrop of our results and similar results of other studies (e.g., Zahedi & van Eck, 2014), we saw the necessity of normalizing reader counts with respect to subject categories and proposed corresponding indicators.

As we have already outlined in section 2, we would like to extend the concept of normalization of altmetrics data (Mendeley data) by focusing the impact measurement on certain sectors of society or status groups. It is the goal of the extended concept that one is able to measure normalized impact on certain groups (e.g., students, professors, or lecturers) and sectors (e.g., educational or research) as aggregated reader groups. However, the target-specific impact measurement only makes sense if there are differences in reader counts between Mendeley status groups.

Table 1. Average reader counts for Mendeley status groups (n=1,388,426 papers)

| Mendeley status group | Number of reader counts | Average reader counts | Percentage of papers with zero reader counts |
|---|---|---|---|
| Student, Bachelor | 1,479,279 | 1.07 | 58.76% |
| Student Master | 2,745,139 | 1.98 | 42.58% |
| Student, post-graduate | 852,265 | 0.61 | 67.50% |
| Student PhD | 4,802,076 | 3.46 | 30.36% |
| Doctoral student | 1,137,291 | 0.82 | 60.83% |
| Lecturer | 251,901 | 0.18 | 86.55% |
| Senior lecturer | 113,757 | 0.08 | 93.05% |
| Associate professor | 909,627 | 0.66 | 65.11% |
| Professor | 611,920 | 0.44 | 73.24% |
| Researcher | 3,191,617 | 2.30 | 38.95% |
| Librarian | 158,430 | 0.11 | 91.21% |
| Other | 696,140 | 0.50 | 72.49% |
| Readers | 16,950,398 | 12.21 | 10.79% |
| Readers with academic status information | 16,949,442 | 12.21 | 10.79% |

Thus, Table 1 shows average reader counts and the percentage of papers with zero reader counts for different Mendeley status groups. For example, there are 158,430 reader counts by librarians; we have 1,388,426 papers in our dataset with 9.79% of papers having at least one librarian's reader. Thus, the average reader count per paper is 0.11. The results in Table 1 reveal significant impact differences between the status groups (which depends on the percentage of papers with zero reader counts): The range is from 0.08 average reader counts for senior lecturers to 3.46 for PhD students. Such readership differences between Mendeley status groups which have also considered field specific differences have been also observed by Zahedi and van Eck (2014): "It is interesting to see that the readership activity of PhD students and professors shows an almost opposite pattern. PhD students have relatively more attention for engineering and less attention for the medical sciences. For professors it is the other way around. Furthermore, professors seem to have a relatively strong focus on mathematics and statistics" (see also Haustein & Larivière, 2014a; Zahedi, Costas, & Wouters, 2014a).

In Table A1 in the appendix, we additionally present the average reader counts for the Mendeley status group "Bachelor students" broken down by WoS subject categories. The table demonstrates that there are not only differences between status groups in reader impact over all WoS subject categories (as shown in Table 1), but also impact differences for papers (articles and reviews) published in different subject categories. Since we focus in the following on bachelor students, the field-specific impact differences are illustrated based on this group (of course, these differences are also visible for all other groups). For example, whereas the average reader count of articles in the WoS subject category "Biodiversity conservation" is 2.51, it is significantly lower in "Biology" with 1.55.

The observed differences in reader impact between status groups justify the proposal in this study to normalize reader impact not only with respect to subject categories, but also to status groups by measuring the field-specific impact target-oriented. The definition of the proposed target-oriented normalized indicator is explained in the next section 5.2.

## 5.2 Definition of target-oriented MNRS

For normalization of citations in bibliometrics, the citation impact of a certain paper is compared with the expected number of citations. The expected impact is the average impact of papers published in the same subject category and publication year as the focal paper. These papers determining the expected value constitute the reference set. The ratio of the citations of the focal paper and its expected citations yields the Normalized Citation Score (NCS). Today, the NCS is the most frequently used method in bibliometrics to normalize citation impact. A NCS of 1 for a paper indicates an average citation impact. A NCS of 1.5 means that the citation impact is 50% above the field average in the particular year (Waltman, van Eck, van Leeuwen, Visser, & van Raan, 2011a; Waltman, van Eck, van Leeuwen, Visser, & van Raan, 2011b).

If a focal paper has been categorized into more than one subject category (e.g., by Thomson Reuters in WoS), the average of all subject-specific NCS values is used. The multiple assignment of papers to subject categories leads to an average NCS of a focal paper which differs from 1. To avoid this deviation, fractional counting (Waltman et al., 2011b), multiplicative counting (Herranz & Ruiz-Castillo, 2012), or full counting with a scaling of all NCS values (Haunschild & Bornmann, 2016b) can be applied. For the Mendeley data used in this study, we decided to apply the multiplicative counting method, although the other counting methods could have been also used. We do not expect completely different conclusions if other counting methods were to be used.

If papers are aggregated to paper sets in order to calculate the normalized impact of a researcher or university, the average of the NCS values in the set is calculated. This average value is called "Mean Normalized Citation Score" (MNCS). Following the definition of the MNCS, we proposed the "Mean Normalized Reader Score" (MNRS) (Haunschild & Bornmann, 2016a). Our normalization procedure for Mendeley data starts with the calculation of the average number of readers per paper ($\rho_c$) in every WoS category (c):

$$\rho_c = \frac{1}{N_c} \sum_{i=1}^{N_c} R_{ic}$$

$R_{ic}$ denotes the raw Mendeley reader count of paper *i*, which has been assigned to WoS category *c*, and $N_c$ is the number of articles or reviews, respectively, assigned to the category (see Table A1 in the appendix). Then, the Mendeley reader count is divided by the average reader number per paper in WoS category *c* ($\rho_c$). This yields the normalized reader score (NRS) for paper *i* in subject category *c*:

$$NRS_{ic} = \frac{R_{ic}}{\rho_c}$$

Problems in the calculation of $NRS_{ic}$ can arise when a subject category has too few reader counts on average (Haunschild, Schier, & Bornmann, 2016). Therefore, we calculate an $NRS_{ic}$ value only if the average reader count in subject category c ($\rho_c$) is at least one. Papers in subject categories with $\rho_c < 1$ have no impact value in this subject category.

Since we use the multiplicative counting method, the average over all NRS values equals exactly one. For specific aggregation levels (e.g., researcher or institute), the overall reader impact can be analyzed by calculating the mean NRS (MNRS). Whereas we used all Mendeley readers of a paper to calculate the NRS previously (Haunschild & Bornmann, 2016a), we propose in this study to use the data of specific user groups to measure impact target specific. This does not change the outlined normalization procedure, but restricts the data analysis to a subset of all readers. As an illustrative example, we show a step by step calculation of the NRS for the article with DOI 10.1016/j.psym.2013.05.004. We recorded a total reader count of 10 for this article. Broken down into academic status groups we have: 3 bachelor students, 2 master students, 3 researchers, 1 librarian, and 1 other reader count. We discard all reader counts except the one from bachelor students. Thus, $R_{ic} = 3$. The article is classified in the WoS subject categories "Psychiatry" (ve) and "Psychology" (vi).

The average bachelor student reader counts for articles in these WoS subject categories are 2.167 and 2.162, respectively. Therefore, we obtain $NRS_{ve} = 1.38$ and $NRS_{vi} = 1.39$. This paper has a reader impact slightly above average in both categories regarding bachelor students. Using the multiplicative counting method, papers assigned to multiple categories do not have a single impact value. For example, if this paper belongs to the publication set of a country and the MNRS is calculated for this set, the paper is not considered only once but three times (with potentially different impact values in each subject category) in the calculation of the average NRS value.

In the theoretical analysis of normalized indicators, one is interested whether the indicator satisfies several properties as desirable conditions for proper impact measurements (Waltman et al., 2011b). For example, an indicator of average impact is said to be consistent if the ranks of two paper sets of equal size (e.g., of two different journals) do not change if both paper sets are expanded by an additional paper with the same citation counts in the same subject categories. Each new indicator should be confronted with these properties. However, since the (target-oriented) MNRS differs only in the data source from the MNCS (and not in the definition), it is not necessary to study the (target-oriented) MNRS theoretically. According to Waltman et al. (2011b), the MNCS satisfies the proposed properties. Therefore, the MNRS also satisfies these properties.

If Mendeley data are used to measure impact target oriented, one can analyze the data on three levels – as proposed by Haustein and Larivière (2014a) (see Table 2): The first level is the Mendeley status group which is provided by Mendeley itself. Each user of Mendeley is asked to assign him- or herself to one status group. Thus, Mendeley reader data for the status groups can be used to calculate normalized reader scores. Following the proposal of Haustein and Larivière (2014a), the status groups can be aggregated on a second level to user types and on a third level to sectors. Depending on the purpose of a study which is intended to measure a specific kind of broader impact, the normalization procedure can be carried out on all three levels including various groups. Recently, Mendeley remapped the academic status groups, e.g., the status group "Assistant Professor" was merged with another status group and the status groups "Post Doc", "Researcher (at a non-Academic Institution)", and "Researcher (at an Academic Institution)" were merged into the new status group "Researcher".

In this study, we decided to explain the procedure using especially the data for the Mendeley status group "Student (Bachelor"). Furthermore, we propose to distinguish the sector "Educational" between receiving and donating education. The Mendeley status group "Student (Bachelor)" is the purest group receiving education. In the following, we show also results for the sector "Education donating" (ED). The sector ED is composed of "Researcher", "Associate Professor", "Lecturer", "Professor", and "Senior Lecturer". The sector ED cannot be pure as most academic status groups are involved in more than one sector, e.g., professors are involved in research and education.

Table 2. Assignment of Mendeley status groups to user types and sectors by Haustein and Larivière (2014a). Note that Mendeley status groups have been changed in early 2016.

| Mendeley status group | User type | Sector |
|---|---|---|
| Assistant Professor | Assistant Professor | Scientific |
| Associate Professor | Associate Professor | Scientific |
| Doctoral Student | PhD Student | Scientific |
| Lecturer | Assistant Professor | Educational |
| Librarian | Librarian | Professional |
| Other Professional | Other Professional | Professional |
| Ph.D. Student | PhD Student | Scientific |
| Post Doc | Postdoc | Scientific |
| Professor | Professor | Scientific |
| Researcher (at a non-Academic Institution) | Researcher (Non Academic) | Professional |
| Researcher (at an Academic Institution) | Researcher (Academic) | Scientific |
| Senior Lecturer | Associate Professor | Educational |
| Student (Bachelor) | Student (Bachelor) | Educational |
| Student (Master) | Student (Postgraduate) | Educational |
| Student (Postgraduate) | Student (Postgraduate) | Educational |

**5.3    Validation of target-oriented MNRS**

As described in section 4, we used data from Altmetric which includes not only Mendeley data but also data from F1000Prime (Bornmann, 2014b, 2015b). F1000Prime is a post-publication peer review system in which 5,000 Faculty Members (senior scientists and leading experts in all areas of biology and medicine) "recommend the most important articles, rating them and providing short explanations for their selections" (http://f1000.com/prime). Besides recommendation and rating of papers, Members can label the papers with specific tags: e.g., "Interesting Hypothesis" (the paper presents new models), "New Finding" (the paper presents original data, models or hypotheses), "Novel Drug Target" (suggests new targets for drug discovery) (http://f1000.com/prime/about/whatis/how). Another tag is "Good for teaching" which labels well written key papers in a field.

In this study, we selected for this study all papers in our dataset with at least one tag and separated two groups (see Table 3): The one group concerns of all papers with the tag "Good for teaching" (perhaps besides other tags). All other papers with at least one tag are pooled in the other group, called "Non-Good for teaching." For both groups, we calculated the target-oriented $MNRS_{BS}$ for the status group "Student (Bachelor"), the $MNRS_{ED}$ for the educational donating sector, and the $MNRS_{ALL}$ based on the reader counts of all other status groups (see Table 3). We selected the group of bachelor students, because only this selection makes sure that reader impact on teaching is measured. The problem with all other groups (e.g., "Associate Professor") is that they are active not only in teaching but also in research. For comparison, we additionally included in Table 3 the MNCS – the mean normalized citation scores of the papers (measuring impact on research). The results in the table show exceptional high reader and citation scores. The papers selected by the Faculty Members for post-publication peer review produce significantly more reader and citation impact than an average paper in the subject categories. This is in accordance with the results of Waltman and Costas (2014) with respect to citation impact.

Table 3. $MNRS_{BS}$, $MNRS_{ED}$, $MNRS_{ALL}$, and MNCS for papers tagged with "Good for teaching" or without this tag

| Tagged group | $MNRS_{BS}$ | $MNRS_{ED}$ | $MNRS_{ALL}$ | MNCS |
|---|---|---|---|---|
| Good for teaching (n=385) | 9.24 | 9.56 | 8.89 | 9.78 |
| Non-Good for teaching (n=904) | 7.69 | 8.41 | 7.81 | 9.34 |
| Difference | 1.55 | 1.15 | 1.08 | 0.44 |
| Total (n=1,289) | 8.91 | 9.38 | 8.71 | 9.85 |

All comparisons in Table 3 show that papers with the tag "Good for teaching" achieve a higher impact than the "Non-Good for teaching" papers. This result is reasonable, because the papers are – according to the assessments of the Faculty Members – well written key papers in a field. The reader and citation impact differences between both groups in Table 3 are in the expected direction. With a value of 1.55, we see the greatest difference for $MNRS_{BS}$. Since bachelor students are a group of academics, which are concerned by teaching in higher education, the "Good for teaching" tag separates "Good for teaching" and "Non-Good for teaching" papers best. A smaller difference in reader impact than for $MNRS_{BS}$ is expectedly shown for $MNRS_{ED}$ (1.15) and $MNRS_{ALL}$ (1.08). The difference decreased to a negligible value in case of the MNCS. For researchers, other criteria than well written key papers seem to be more important (e.g. whether a paper contains interesting new findings or not).

**5.4  Target-oriented reader impact of journals, countries and German institutions**

In this section, we would like to present some results which demonstrate how reader counts can be used to measure target-oriented impact and how the results can be interpreted. We show the impact of journals', countries', and German research institutions' research on bachelor students ($MNRS_{BS}$) and the sector "Education donating" ($MNRS_{ED}$).

Table 4. Twenty journals with the highest $MNRS_{BS}$ (including journals with more than 100 papers – distinct counted – in the dataset and decreasingly sorted by $MNRS_{BS}$). The $MNRS_{ED}$ is added for comparison.

| Journal | Number of papers (multiplicative counting) | Number of papers (distinct counting) | $MNRS_{BS}$ | $MNRS_{ED}$ |
| --- | --- | --- | --- | --- |
| *Nature Biotechnology* | 111 | 111 | 13.97 | 15.52 |
| *Cell* | 872 | 436 | 9.63 | 10.34 |
| *Nature Methods* | 156 | 156 | 9.32 | 10.34 |
| *Nature* | 862 | 862 | 8.36 | 9.20 |
| *Lancet* | 224 | 224 | 6.65 | 7.35 |
| *American Economic Review* | 241 | 241 | 6.41 | 3.74 |
| *Nature Neuroscience* | 220 | 220 | 6.39 | 8.04 |

| Journal | | | | |
|---|---|---|---|---|
| Nature Medicine | 465 | 155 | 5.67 | 8.50 |
| New England Journal of Medicine | 353 | 353 | 5.58 | 8.11 |
| Cell Stem Cell | 224 | 112 | 5.32 | 5.98 |
| Science | 826 | 826 | 5.21 | 6.30 |
| Immunity | 149 | 149 | 5.13 | 6.71 |
| Nature Immunology | 120 | 120 | 4.92 | 7.24 |
| Nature Genetics | 192 | 192 | 4.76 | 6.90 |
| Nature Protocols | 197 | 197 | 4.73 | 4.02 |
| Nature Chemical Biology | 139 | 139 | 4.62 | 4.37 |
| Neuron | 400 | 400 | 4.55 | 5.65 |
| Computers & Education | 220 | 217 | 4.24 | 3.32 |
| Decision Support Systems | 366 | 183 | 4.02 | 2.73 |
| Genome Research | 561 | 187 | 3.97 | 5.89 |

Table 4 shows the twenty journals with the highest $MNRS_{BS}$ values and the corresponding $MNRS_{ED}$ values out of 7,948 journals covered by the WoS. Against the backdrop that MNRS=1 identifies journals with an average reader impact on bachelor students/ in the education donating sector in the corresponding subject categories, the impact of the journals in the table is extraordinarily high. For example, *Genome Research* as the journal with the lowest impact on bachelor students in Table 4 publishes papers which have on average four times as many readers among Bachelor students than can be expected in the corresponding subject areas. The three journals in the table with the highest impact on these students and on the educational donating sector in general are *Nature Biotechnology* ($MNRS_{BS}$=13.97, $MNRS_{ED}$ = 15.52), *Cell* ($MNRS_{BS}$=9.63, $MNRS_{ED}$ = 10.34), and *Nature Methods* ($MNRS_{BS}$=9.32, $MNRS_{ED}$ = 10.34).

Table 5. Countries with the highest $MNRS_{BS}$ (including countries with at least 10,000 papers – distinct counted – in the dataset and decreasingly sorted by $MNRS_{BS}$). The $MNRS_{ED}$ is added for comparison.

| Country | Number of papers (multiplicative counting) | Number of papers (distinct counting) | $MNRS_{BS}$ | $MNRS_{ED}$ |
|---|---|---|---|---|
| Great Britain | 230,152 | 64,354 | 1.45 | 1.56 |

| Denmark | 36,500 | 10,185 | 1.37 | 1.41 |
| Switzerland | 49,895 | 15,108 | 1.34 | 1.55 |
| Netherlands | 98,294 | 23,745 | 1.31 | 1.42 |
| United States | 1,089,729 | 224,502 | 1.31 | 1.46 |
| Australia | 147,985 | 35,399 | 1.28 | 1.24 |
| Sweden | 51,281 | 14,823 | 1.27 | 1.33 |
| Belgium | 40,831 | 11,469 | 1.25 | 1.33 |
| Canada | 157,419 | 37,750 | 1.22 | 1.24 |
| Germany | 200,885 | 48,971 | 1.19 | 1.35 |
| Spain | 116,083 | 28,327 | 1.13 | 1.19 |
| Brazil | 88,678 | 19,909 | 1.12 | 0.79 |
| France | 142,577 | 31,291 | 1.04 | 1.27 |
| Italy | 127,709 | 30,027 | 1.02 | 1.07 |
| Japan | 137,904 | 31,201 | 0.84 | 0.87 |
| South Korea | 99,053 | 22,108 | 0.75 | 0.67 |
| Taiwan | 73,273 | 12,428 | 0.73 | 0.63 |
| India | 65,005 | 18,739 | 0.71 | 0.65 |
| China | 377,451 | 93,879 | 0.64 | 0.61 |

In the second analysis, we investigated the target-oriented reader impact of papers published by countries worldwide. The results are presented in Table 5. In comparison to Table 4, we observe significantly lower $MNRS_{BS}$ and $MNRS_{BS}$. Also, we can differentiate between countries producing papers with more or less impact on bachelor students/ in the education donating sector than average. Thus, the countries are able to produce more or less useful papers for these students – on average and across subject categories. Researchers from Great Britain are able to publish papers with the higher target-oriented impact on bachelor students ($MNRS_{BS}$=1.45) and in the educational donating sector ($MNRS_{ED}$=1.56). On average, their papers produce 45%/ 56% more impact on Bachelor students/ on the educational donating sector than can be expected in the corresponding subject categories. The country with the lowest impact on both groups in Table 5 is China: Researchers from China publish papers which are significantly less useful for Bachelor students (100%-64%=36%) and the educational donating sector (100%-61%=39%) than the average paper in a subject category.

Table 6. Twenty institutions in Germany with the highest $MNRS_{BS}$ (including institutions with at least 1,000 papers – distinct counted – in the dataset and decreasingly sorted by $MNRS_{BS}$). The $MNRS_{ED}$ is added for comparison.

| Institution in Germany | Number of papers (multiplicative counting) | Number of papers (distinct counting) | $MNRS_{BS}$ | $MNRS_{ED}$ |
|---|---|---|---|---|
| Universität Bonn | 3,721 | 1,493 | 1.39 | 1.52 |
| Universität Göttingen | 3,752 | 1,550 | 1.34 | 1.50 |
| Bayerische Julius-Maximilians Universität Würzburg | 2,943 | 1,138 | 1.24 | 1.27 |
| Universität Heidelberg | 6,328 | 2,446 | 1.23 | 1.34 |
| Rheinisch-Westfälische Technische Hochschule (RWTH) | 3,174 | 1,190 | 1.21 | 1.28 |
| LMU München | 6,785 | 2,809 | 1.21 | 1.38 |
| Goethe University Frankfurt | 3,387 | 1,313 | 1.21 | 1.26 |
| Universität Freiburg | 4,626 | 1,583 | 1.20 | 1.46 |
| Charité - Universitätsmedizin Berlin | 5,593 | 2,060 | 1.19 | 1.36 |
| TU München | 4,948 | 2,058 | 1.18 | 1.56 |
| Universität Tübingen | 4,789 | 1,802 | 1.18 | 1.26 |
| Universität Ulm | 2,304 | 1,002 | 1.16 | 1.27 |
| Universität Köln | 3,555 | 1,333 | 1.16 | 1.39 |
| Universität Münster | 3,362 | 1,324 | 1.12 | 1.15 |
| TU Dresden | 3,301 | 1,316 | 1.11 | 1.24 |
| Universität Hamburg | 4,315 | 1,743 | 1.08 | 1.20 |
| Universität Mainz | 2,595 | 1,063 | 1.05 | 1.10 |
| Universität Erlangen-Nürnberg | 3,360 | 1,327 | 1.04 | 1.21 |
| Heinrich-Heine-Universität Düsseldorf | 2,817 | 1,159 | 1.02 | 1.21 |
| Universität Jena | 2,747 | 1,063 | 1.02 | 1.04 |

In the final analysis, we computed the target-oriented reader impact of papers published by German institutions. We focus on Germany, because we only have paper assignments to institutions from this country in the MPDL in-house database (and not from other countries) which are cleaned for address variants and are thus reliably collected for research evaluation purposes. Table 6 shows the twenty institutions in Germany with the highest $MNRS_{BS}$. Also, the $MNRS_{ED}$ is added for comparison. All institutions have impact scores above average; the institutions publishing papers with the highest impact on bachelor students are the universities of Bonn ($MNRS_{BS}$=1.39) and Göttingen ($MNRS_{BS}$=1.34). The university with highest reader impact on the educational donating sector is TU München ($MNRS_{ED}$=1.56).

# 6 Discussion

In scientometrics, citations are the standard data source for measuring the impact of papers on scientists. Thus, citations are a target-oriented impact indicator which is restricted to the scientific sector. Altmetrics have been proposed to measure the impact of papers on other sectors of society or other groups of people than scientists, respectively (Bornmann, 2014a). Altmetrics may reflect wider impacts "because non-academics widely publish online and may sometimes discuss researchers and research-related issues" (Thelwall, Kousha, Dinsmore, & Dolby, 2016). Currently, it is the tendency in analyzing altmetrics data to measure impact on many sectors of society by not differentiating between different target groups (e.g., tweets by science communicators, scientists, or members of the public). This tendency is especially reflected in the efforts to design composite indicators (Williams, 2016), although altmetrics data are especially suitable to measure impact specifically (target-oriented): "Citations do not measure readership and do not account for the impact of scholarly papers on teaching, professional practice, technology development, and nonacademic audiences" (Weller, 2015, p. 264).

In this study, we introduce impact scores which are intended to measure impact of papers target-oriented. Here, we follow proposals, such as those published by Thelwall et al. (2016), to give a clear indication of the type of impact reflected by the used indicator. Our study is based on Mendeley readership data. Since this data is affected by field differences (Haunschild & Bornmann, 2016a) independent of the quality of the publication – as many other altmetrics and bibliometrics data – we propose to apply the field-normalized reader score (MNRS) in a target-oriented manner. Mendeley readership data are especially suitable for the calculation of these reader scores, because Mendeley provides information about which papers were used (read) and by how many Mendeley users (Haustein, 2014). Mendeley provides such readership data broken down by academic status, scientific discipline, and country. The target-oriented indicator can be calculated on different levels: It can be used to measure the normalized impact on different Mendeley status groups (e.g., Bachelor students). However, the data for the groups can also be aggregated to user groups (e.g., professors) and sectors (e.g. education) (Haustein & Larivière, 2014a). However, the aggregation faces the problem that many groups cannot be unequivocally assigned to sectors. For example, professors are not only active in the educational, but also in the research sector.

The target-oriented measurement of impact is not only restricted to Mendeley data. Also, other altmetrics data can be used. For example, Bornmann, Haunschild, and Marx (in press) explore the use of academic paper mentions in policy-related documents as a possible altmetrics source. These paper mentions are provided by Altmetric for impact measurements on the policy sector. Also, Twitter data could be used for target-specific measurements (Bornmann & Haunschild, 2016a). Altmetric assigns Twitter users to specific groups of people (e.g., practitioners, members of the public, researchers, or science communicators).

By field-normalizing altmetrics data (Bornmann & Haunschild, 2016a; Haunschild & Bornmann, 2016a) and measuring impact target-oriented (Bornmann et al., in press), the methods of measuring impact with altmetrics data are developed towards an advanced level – similar to the advanced level in bibliometrics. However, there remain questions in scientometrics concerning the meaning of altmetrics data and their significance for research evaluation purposes: Does a save or read mean the same as a citation (Bar-Ilan et al., 2014; Rodgers & Barbrow, 2013)? According to Rousseau and Ye (2013) "'shares' lack authority and scientific credibility so that the use of altmetrics may still be somewhat premature." For Shema, Bar-Ilan, and Thelwall (2014), "the lack of context makes it difficult to determine the underlying use made of a bookmarked article. The users might be called readers but it is possible that they have not read the item they bookmarked or that they have read it but did not make use of it. On the other hand, it could be that they use reference managers to easily access important articles over and again". In future studies, real readers might be better identified if only those reads are counted where "users make the effort to annotate documents with tags" (Haustein, 2014).

Research on the use of altmetrics in research evaluation is still in an early phase, but is quickly developing (Bornmann & Haunschild, in press; Weller, 2015) – similar to the h index research after the introduction of the new index (Bornmann & Marx, 2012; Hirsch, 2005). According to Haustein, Larivière, Thelwall, Amyot, and Peters (2014) qualitative studies are especially important to explore the motivations and reasons for sharing, saving, tweeting etc.

# 7 Conclusions

We extend in this study the field-normalized reader impact indicator based on Mendeley data (the mean normalized reader score, MNRS) to a target-oriented field-normalized impact indicator. For example, $MNRS_{ED}$ measures reader impact on the sector of educational donation, i.e., teaching. This indicator can show the ability of journals, countries, and academic institutions to publish papers which are below or above the average impact of papers on a specific sector in society. For example, the method allows to measure the impact of scientific papers on PhD students – controlling for the field in which the papers have been published and their publication year.

# Acknowledgements

The bibliometric data used in this paper are from an in-house database developed and maintained by the Max Planck Digital Library (MPDL, Munich) and derived from the Science Citation Index Expanded (SCI-E), Social Sciences Citation Index (SSCI), and Arts and Humanities Citation Index (AHCI) prepared by Thomson Reuters (Philadelphia, Pennsylvania, USA). The Mendeley reader counts were retrieved via the API of Mendeley. The F1000Prime tags were taken from a data set retrieved from Altmetric on December 19, 2015.

# Appendix A

Table A1: Average reader counts for the Mendeley status group "Bachelor students" (BS) broken down by WoS subject category and document type. Average reader counts between 0 and 1 are rounded down to zero. As outlined in section 5.2, we calculate an $NRS_{ic}$ value only if the average readers score in a subject category is at least one. Papers in subject categories with lower impact than 1 have no impact value.

| WoS subject category | Articles | Reviews |
|---|---|---|
| Audiology & speech-language pathology | 0.00 | 3.91 |
| Acoustics | 0.00 | 1.32 |
| Agricultural economics & policy | 1.45 | 1.17 |
| Agricultural engineering | 1.96 | 7.65 |
| Agriculture, dairy & animal science | 0.00 | 1.67 |
| Agriculture, multidisciplinary | 0.00 | 2.04 |
| Agronomy | 0.00 | 2.54 |
| Allergy | 1.02 | 1.80 |
| Anatomy & morphology | 0.00 | 2.43 |
| Andrology | 0.00 | 2.01 |
| Anesthesiology | 0.00 | 1.33 |
| Anthropology | 1.20 | 1.08 |
| Archaeology | 0.00 | 1.19 |
| Area studies | 0.00 | 0.00 |
| Art | 0.00 | 0.00 |
| Astronomy & astrophysics | 0.00 | 0.00 |
| Automation & control systems | 0.00 | 1.18 |
| Behavioral sciences | 2.48 | 7.18 |
| Biochemical research methods | 1.24 | 3.20 |
| Biochemistry & molecular biology | 1.37 | 3.87 |
| Biodiversity conservation | 2.51 | 6.10 |
| Biology | 1.55 | 4.13 |
| Biophysics | 0.00 | 4.04 |
| Biotechnology & applied microbiology | 1.38 | 4.45 |
| Business | 3.23 | 4.37 |
| Business, finance | 1.66 | 1.82 |
| Cell & tissue engineering | 1.63 | 4.28 |
| Cultural studies | 0.00 | 1.00 |
| Cardiac & cardiovascular system | 0.00 | 2.03 |
| Cell biology | 1.82 | 4.90 |
| Chemistry, analytical | 0.00 | 2.31 |
| Chemistry, applied | 0.00 | 3.06 |
| Chemistry, inorganic & nuclear | 0.00 | 1.21 |
| Chemistry, medicinal | 0.00 | 2.05 |
| Chemistry, multidisciplinary | 0.00 | 2.87 |
| Chemistry, organic | 0.00 | 2.00 |
| Chemistry, physical | 0.00 | 2.62 |
| Clinical neurology | 1.21 | 2.52 |
| Communication | 1.72 | 2.00 |
| Computer science, artificial intelligence | 0.00 | 2.13 |
| Computer science, cybernetics | 1.00 | 1.82 |
| Computer science, hardware & architecture | 0.00 | 4.20 |
| Computer science, information systems | 1.01 | 3.76 |
| Computer science, interdisciplinary applications | 0.00 | 2.65 |
| Computer science, software engineering | 0.00 | 4.35 |
| Computer science, theory & methods | 0.00 | 2.12 |
| Construction & building technology | 1.11 | 4.48 |
| Criminology & penology | 1.43 | 2.66 |
| Critical care medicine | 1.34 | 1.65 |
| Crystallography | 0.00 | 0.00 |
| Demography | 0.00 | 4.00 |
| Dentistry, oral surgery & medicine | 0.00 | 1.56 |
| Dermatology | 0.00 | 1.66 |
| Developmental biology | 1.56 | 4.03 |
| Ecology | 2.33 | 6.85 |
| Economics | 1.61 | 1.66 |
| Education & educational research | 1.16 | 2.00 |
| Education, scientific disciplines | 1.35 | 2.52 |
| Education, special | 1.96 | 5.03 |
| Electrochemistry | 0.00 | 3.45 |
| Emergency medicine | 1.09 | 1.60 |
| Endocrinology & metabolism | 1.33 | 2.93 |
| Energy & fuels | 1.24 | 4.87 |
| Engineering, aerospace | 0.00 | 3.16 |
| Engineering, biomedical | 1.07 | 4.02 |
| Engineering, chemical | 0.00 | 3.21 |
| Engineering, civil | 0.00 | 3.58 |
| Engineering, electrical & electronic | 0.00 | 1.98 |
| Engineering, environmental | 1.34 | 5.41 |
| Engineering, geological | 0.00 | 1.20 |
| Engineering, industrial | 1.83 | 4.32 |
| Engineering, manufacturing | 0.00 | 2.25 |
| Engineering, marine | 0.00 | 1.88 |
| Engineering, mechanical | 0.00 | 1.73 |
| Engineering, multidisciplinary | 0.00 | 1.25 |
| Engineering, ocean | 0.00 | 1.50 |
| Entomology | 0.00 | 2.60 |
| Environmental sciences | 1.35 | 4.02 |
| Environmental studies | 2.04 | 3.24 |
| Ergonomics | 2.32 | 4.48 |
| Ethics | 1.51 | 1.29 |
| Ethnic studies | 0.00 | 1.09 |
| Evolutionary biology | 2.38 | 6.46 |
| Family studies | 1.93 | 2.19 |
| Fisheries | 0.00 | 2.57 |
| Food science & technology | 1.08 | 3.56 |
| Forestry | 0.00 | 2.02 |
| Gastroenterology & hepatology | 0.00 | 1.67 |
| Genetics & heredity | 1.46 | 5.30 |
| Geochemistry & geophysics | 0.00 | 1.75 |
| Geography | 1.46 | 2.38 |
| Geography, physical | 1.17 | 3.10 |
| Geology | 0.00 | 0.00 |
| Geosciences, multidisciplinary | 0.00 | 2.28 |
| Geriatrics & gerontology | 1.73 | 4.11 |
| Gerontology | 1.55 | 2.84 |
| Health care sciences & services | 1.31 | 1.92 |
| Health policy & services | 1.27 | 1.55 |
| Hematology | 0.00 | 2.00 |
| History & Philosophy of Science | 0.00 | 1.38 |
| History of Social Sciences | 0.00 | 0.00 |
| Horticulture | 0.00 | 2.13 |
| Hospitality, leisure, sport & tourism | 2.38 | 4.90 |
| Imaging science & photographic technology | 0.00 | 3.00 |
| Immunology | 1.10 | 3.34 |
| Industrial relations & labor | 1.22 | 1.00 |
| Infectious diseases | 1.17 | 2.68 |
| Information science & library science | 1.86 | 3.05 |
| Instruments & instrumentation | 0.00 | 2.82 |
| Integrative & complementary medicine | 1.91 | 3.57 |
| International relations | 1.39 | 1.52 |
| Language & linguistics theory | 0.00 | 0.00 |
| Law | 0.00 | 0.00 |
| Limnology | 0.00 | 1.50 |
| Linguistics | 0.00 | 2.10 |
| Management | 2.60 | 3.69 |
| Marine & freshwater biology | 1.38 | 3.84 |
| Materials science, biomaterials | 1.07 | 4.78 |
| Materials science, ceramics | 0.00 | 0.00 |
| Materials science, characterization, testing | 0.00 | 0.00 |
| Materials science, coatings & films | 0.00 | 1.31 |
| Materials science, composites | 0.00 | 2.72 |
| Materials science, multidisciplinary | 0.00 | 2.69 |
| Materials science, paper & wood | 0.00 | 0.00 |
| Materials science, textiles | 0.00 | 2.00 |
| Mathematical & computational biology | 1.21 | 2.46 |
| Mathematics, interdisciplinary applications | 0.00 | 0.00 |
| Mechanics | 0.00 | 1.69 |
| Medical ethics | 1.29 | 1.44 |
| Medical informatics | 1.05 | 3.25 |
| Medical laboratory technology | 0.00 | 1.85 |
| Medicine, general & internal | 1.38 | 2.90 |
| Medicine, legal | 1.28 | 1.63 |
| Medicine, research & experimental | 1.07 | 2.83 |
| Metallurgy & metallurgical engineering | 0.00 | 0.00 |
| Meteorology & atmospheric sciences | 0.00 | 1.78 |
| Microbiology | 1.32 | 4.73 |
| Microscopy | 0.00 | 1.49 |
| Mineralogy | 0.00 | 1.34 |
| Mining & mineral processing | 0.00 | 0.00 |
| Multidisciplinary sciences | 2.01 | 4.99 |
| Music | 0.00 | 0.00 |
| Mycology | 0.00 | 2.42 |
| Nanoscience & nanotechnology | 0.00 | 2.42 |
| Neuroimaging | 1.89 | 4.26 |
| Neurosciences | 1.97 | 5.37 |
| Nuclear science & technology | 0.00 | 0.00 |
| Nursing | 1.72 | 2.68 |
| Nutrition & dietetics | 2.11 | 4.62 |
| Obstetrics & gynecology | 0.00 | 1.77 |
| Oceanography | 0.00 | 2.52 |
| Oncology | 0.00 | 2.04 |
| Operations research & management science | 1.35 | 3.93 |
| Ophthalmology | 0.00 | 1.26 |
| Optics | 0.00 | 2.18 |
| Ornithology | 1.24 | 2.44 |
| Orthopedics | 1.17 | 2.32 |
| Otorhinolaryngology | 0.00 | 1.16 |
| Primary health care | 1.50 | 2.83 |
| Paleontology | 0.00 | 1.12 |
| Parasitiology | 1.49 | 3.83 |
| Pathology | 0.00 | 2.03 |
| Pediatrics | 1.07 | 1.76 |
| Peripheral vascular diseases | 0.00 | 1.75 |
| Pharmacology & pharmacy | 0.00 | 2.92 |
| Philosophy | 0.00 | 0.00 |
| Physics, applied | 0.00 | 2.08 |
| Physics, atomic, molecular & chemical | 0.00 | 1.30 |
| Physics, condensed matter | 0.00 | 2.06 |
| Physics, fluids & plasmas | 0.00 | 1.77 |
| Physics, mathematical | 0.00 | 0.00 |
| Physics, multidisciplinary | 0.00 | 1.07 |
| Physics, nuclear | 0.00 | 0.00 |
| Physics, particles & fields | 0.00 | 0.00 |
| Physiology | 1.23 | 3.18 |
| Planning & development | 1.75 | 1.64 |
| Plant sciences | 0.00 | 3.75 |
| Political science | 1.08 | 1.03 |
| Polymer science | 0.00 | 3.24 |
| Psychiatry | 2.17 | 3.90 |
| Psychology | 2.16 | 5.35 |
| Psychology, applied | 2.67 | 4.82 |
| Psychology, biological | 2.48 | 6.98 |
| Psychology, clinical | 2.54 | 4.50 |
| Psychology, developmental | 2.68 | 4.83 |
| Psychology, educational | 1.76 | 2.67 |
| Psychology, experimental | 2.66 | 7.08 |
| Psychology, mathematical | 1.24 | 2.65 |
| Psychology, multidisciplinary | 2.63 | 5.52 |
| Psychology, psychoanalysis | 0.00 | 0.00 |
| Psychology, social | 3.00 | 4.02 |
| Public administration | 0.00 | 0.00 |
| Public, environmental & occupational health | 1.40 | 2.62 |
| Radiology, nuclear medicine & medical imaging | 0.00 | 1.54 |
| Rehabilitation | 2.03 | 3.85 |
| Remote sensing | 0.00 | 2.67 |
| Reproductive biology | 0.00 | 2.35 |
| Respiratory system | 0.00 | 1.50 |
| Rheumatology | 1.18 | 2.16 |
| Robotics | 1.25 | 2.72 |
| Social issues | 1.09 | 2.08 |
| Social sciences, biomedical | 1.74 | 2.84 |
| Social sciences, interdisciplinary | 1.36 | 2.84 |
| Social sciences, mathematical methods | 0.00 | 0.00 |
| Social work | 1.49 | 2.11 |
| Sociology | 1.27 | 1.43 |
| Soil science | 1.01 | 3.77 |
| Spectroscopy | 0.00 | 1.00 |
| Sport sciences | 2.55 | 5.58 |
| Statistics & probability | 0.00 | 0.00 |
| Substance abuse | 1.69 | 3.28 |
| Surgery | 0.00 | 1.18 |
| Telecommunications | 0.00 | 2.47 |
| Thermodynamics | 0.00 | 3.34 |
| Toxicology | 1.01 | 2.41 |
| Transplantation | 0.00 | 1.23 |
| Transportation | 1.94 | 4.13 |
| Transportation science & technology | 0.00 | 1.37 |
| Tropical medicine | 1.64 | 3.88 |
| Urban studies | 1.57 | 1.96 |
| Urology & nephrology | 0.00 | 1.13 |
| Veterinary sciences | 0.00 | 2.26 |
| Virology | 1.28 | 3.23 |
| Water resources | 0.00 | 2.67 |
| Women's studies | 1.26 | 2.88 |
| Zoology | 1.24 | 2.59 |